\documentclass[fleqn,10pt]{wlscirep}
\usepackage[utf8]{inputenc}
\usepackage[T1]{fontenc}
\usepackage{subfig}
\usepackage{csquotes}
\usepackage{booktabs}
\usepackage{arydshln}
\usepackage{gensymb}

\title{A unified repository for pre-processed climate data weighted by gridded economic activity}

\author[1]{Marco Gortan}
\author[2,3]{Lorenzo Testa}
\author[3,*]{Giorgio Fagiolo}
\author[3,4,*]{Francesco Lamperti}

\affil[1]{School of Finance, University of St. Gallen, St. Gallen, Switzerland}
\affil[2]{Department of Statistics and Data Science, Carnegie Mellon University, Pittsburgh PA, US}
\affil[3]{Institute of Economics and L'EMbeDS, Sant'Anna School of Advanced Studies, Pisa, Italy}
\affil[4]{RFF-CMCC European Institute on Economics and the Environment, Milan, Italy}

\affil[*]{Co-corresponding authors: Francesco Lamperti (f.lamperti@santannapisa.it); Giorgio Fagiolo (g.fagiolo@santannapisa.it)}

\begin{abstract}
Although high-resolution gridded climate variables are provided by multiple sources, the need for country and region-specific climate data \textit{weighted} by indicators of economic activity is becoming increasingly common in environmental and economic research. We process available information from different climate data sources to provide spatially aggregated data with global coverage for both countries (GADM0 resolution) and regions (GADM1 resolution) and for a variety of climate indicators (average precipitations, average temperatures, average SPEI). 
We weigh gridded climate data by population density or by night light intensity -- both proxies of economic activity -- before aggregation.
Climate variables are measured daily, monthly, and annually, covering (depending on the data source) a time window from 1900 (at the earliest) to 2023. 
We pipeline all the preprocessing procedures in a unified framework, which we share in the open-access \textit{Weighted Climate Data Repository} web app. Finally, we validate our data through a systematic comparison with those employed in leading climate impact studies.

\vspace{1cm}

\textbf{Keywords}: climate and weather data; spatial weighting; impact assessment; climate econometrics.  
\end{abstract}
\begin{document}

\flushbottom
\maketitle

\thispagestyle{empty}

\section*{Background \& Summary}
Climate change and weather events have been shown to adversely affect a wide spectrum of natural and socio-economic activities \cite{dell2014we,carleton2016social}. A blossoming body of literature reports evidence of significant and non-linear impacts on agricultural \cite{schlenker2009nonlinear} and economic production \cite{burke2015global,kalkuhl2020impact,kotz2022effect}, conflict \cite{abel2019climate}, income inequality \cite{palagi2022climate}, mortality \cite{carleton2022}, energy consumption \cite{auffhammer2014measuring}, and the list is far from being conclusive. Most of these studies test the presence of a significant statistical association between climate variables and socioeconomic indicators, adopting either cross-section or panel-data approaches \cite{hsiang2016climate, auffhammer2018quantifying}. 

One common challenge is that weather data are typically available at a much finer spatiotemporal resolution than socioeconomic variables. While indicators such as industrial production, GDP, employment, and fatalities are typically collected annually -- at region or country breakdowns -- temperatures, precipitations, and other weather variables are instead available at gridded levels and daily frequency. Hence, the common approach requires weather-related variables to be aggregated to match lower temporal frequencies and the geographical boundaries of administrative units. 

This process is not straightforward and often requires the use of weights proxying the geographical distribution of economic activities. Indeed, when studying the impact of climatic conditions and weather events on the economy, it is crucial to account for the different exposure of socioeconomic activities within an administrative region. For example, average temperatures in the Mojave Desert (California, US) during the summer may be way higher than in Los Angeles (California, US), but the size of economic activities in the two locations is not even comparable. Indeed, one may easily argue that labor productivity in California is much more affected by temperatures in Los Angeles than in desert areas. Thus, a simple aggregation of climate data that does not account for the geography of socioeconomic activities could introduce a bias in the evaluation of climate impacts, especially when the variability across administrative regions is central to the identification of the effect \cite{hsiang2016climate,auffhammer2018quantifying}. Further, when a weather-related phenomenon occurs at regional level, in response to averaged weather, the weighting scheme is crucial to reflect the relative overall importance of weather in different regions. For instance, weighting rainfall by the distance from ashore could help to predict the declaration of states of emergency. 

Spatially weighted data are increasingly employed in the literature exploring the impacts of climate change and weather events on socio-economic activities. For example, Burke et al.\cite{burke2015global}, in a seminal study assessing the effect of global warming on the dynamics of economic production, employ population-weighted temperatures and precipitations to measure gradual climate change. Accordingly, a number of studies have been relying on Burke et al. dataset to explore the impact of climate on economic inequality and growth \cite{palagi2022climate,diffenbaugh2019global,alessandri2021macroeconomic}. Furthermore, population weighting is not limited to the case of average temperatures and total precipitations, as it is increasingly employed for a variety of additional climate indicators, e.g., in the evaluation of heating and cooling degree days \cite{spinoni2021global}.

However, replicating published studies using spatially weighted climate data is difficult, as the exact procedure employed to obtain weighted climate variables used for impact assessment is often unclear, under-discussed, or not reported at all in existing contributions. This poses a potential problem, as the way in which weighting is performed may depend on a number of different key factors and choices\cite{wei2022comparison}. Among them, the sources of data used for the construction of weights, the adjustments employed to align gridded information to the borders of administrative regions, and the eventual use of a base year are all elements that can sensibly affect the construction of spatially weighted climate indicators. This also undermines exercises trying to employ existing datasets containing spatially-weighted climate variables (e.g., made available in online repositories as supplementary material of published papers) in further studies or analyses. Indeed, in absence of clear guidelines and documentation, it becomes very hard to build homogenized datasets covering different sets of countries or regions and longer time series (i.e., more recent years). 

Here, we argue that the lack of a harmonized, documented, cross-validated, and open-access repository for climate variables that are spatially weighted by economic activity hinders a rigorous and robust estimation of the social and economic impacts of climate change. This may partly explain why unweighted climate indicators are still employed in several studies. For example, Kotz et al.\cite{kotz2022effect} construct a number of indicators proxying the yearly distribution of rainfall within national and subnational regions without accounting for the spatial distribution of economic activities, and use such indicators to show the adverse impact of precipitation extremes on economic growth. Furthermore, spatially unweighted climate data are also employed in the emergent macro-econometric literature on climate impacts\cite{ponticelli2023temperature,donadelli2022temperature,cipollini2023temperature,donadelli2021global}.

In this paper, we try to close this gap by introducing a unified repository that pipelines the preprocessing and weighting procedures of gridded climate data into a documented, intuitive, and open-access interface. The repository allows researchers to get ready-to-use climate variable datasets aggregated at national and sub-national levels, with global coverage over the period 1900--2023. In particular, the \textit{Weighted Climate Data Repository} provides a user-friendly dashboard to explore and download key climate variables under customizable weighting schemes, temporal frequency, timeframe, administrative level, and file format. 

Our repository is intended to support the climate impact assessment community, which is constantly enlarging and increasingly opening to scientists and researchers who aim to work with datasets compiled at the administrative level (e.g., economists and public policy scholars). Indeed, by offering a unified and harmonized access to a wealth of publicly available yet dispersed and unweighted climate and weather indicators, we aim to improve the replicability of impact assessment studies, increase the transparency of data management practices, and incentivize the community to test the robustness of estimates to the choice of data sources and aggregation strategies. The \textit{Weighted Climate Data Repository} is available and maintained at \url{https://weightedclimatedata.streamlit.app}.

\section*{Methods}
\begin{figure}
    \centering
    \includegraphics[width=\textwidth]{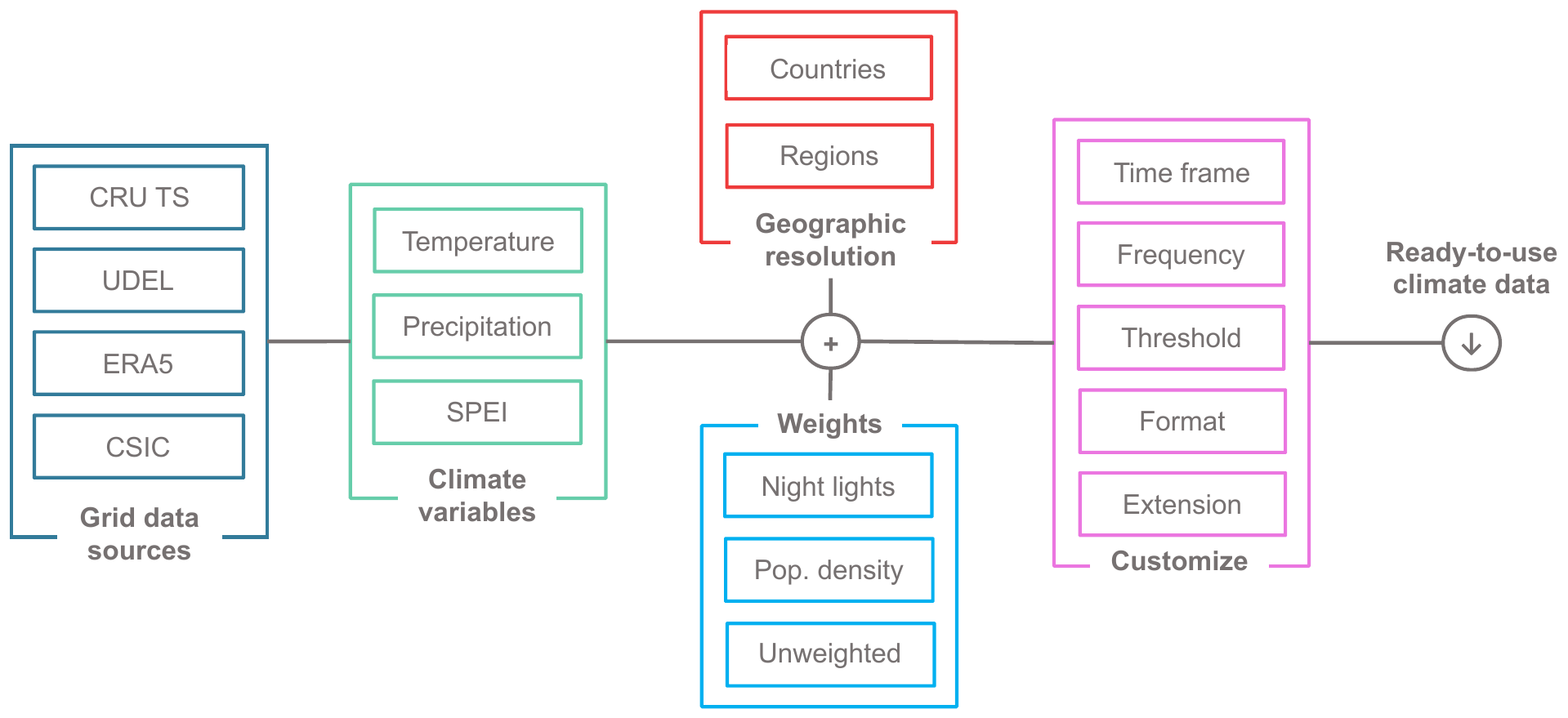}
    \caption{The \textit{Weighted Climate Data Repository} workflow. Users can combine gridded climate variables, gridded indicators of economic activity, and administrative boundaries to achieve regional climate variables \textit{weighted} by economic activity.}
    \label{fig:pipeline}
\end{figure}

The logical steps behind the construction of our repository are illustrated in Figure \ref{fig:pipeline}. We combine different datasets, including gridded climate variables from multiple open-access sources, gridded indicators of spatial socio-economic activity, and administrative boundaries at different levels of resolution. The main objective is to obtain climate data that are weighted by socio-economic indicators according to different strategies that are customizable by the user. To achieve this, our procedure follows three key steps:

\begin{itemize}
    \item \textbf{Selection}: In the first step, we choose (i) a specific set of gridded climate variables of interest, (ii) the desired geographical resolution, and (iii) a gridded economic activity indicator for constructing the aggregation weights.
    
    \item \textbf{Computation of weights}: Next, we integrate the selected information to derive a gridded weighted version of each climate variable. This process ensures that the socio-economic indicators are appropriately considered in the analysis.

    \item \textbf{Aggregation}: Finally, we aggregate the gridded weighted observations across the regions defined by the chosen geographical resolution. This step allows us to obtain a comprehensive view of climate data at the desired level of granularity.
\end{itemize}

Our repository includes an interactive interface that enables users to customize the aggregation process and the format of downloadable datasets. They can modify parameters such as the base year for constructing weights, the frequency of climate data (i.e., daily, monthly, yearly), and the time span of interest. Additionally, users can query the database to access specific information tailored to their end-use requirements.

\subsection*{Gridded variables and administrative boundaries}
The core of the \textit{Weighted Climate Data Repository} rests on two groups of gridded variables: climate variables and indicators of economic activity. These variables, together with administrative boundaries, serve as the fundamental components of our repository. Table \ref{tab:sources} shows all the sources of data we exploit in our work.

\subsubsection*{Climate data}
We leverage raw gridded climate data from four sources that are routinely used in climate impact studies: Climate Research Unit Time-Series \cite{harris2020version} (CRU TS v4.07, available from 1901 until 2022), Consejo Superior de Investigaciones Científicas \cite{vicente2010new} (CSIC v2.7, 1901--2020),
ECMWF Reanalysis v5 \cite{hersbach2020era5} (ERA5, 1940--2023), and University of Delaware \cite{willmott2000terrestrial} (UDEL v5.01, 1900--2017). CRU TS, UDEL, and CSIC provide data at the grid resolution of $0.5^{\circ}\times0.5^{\circ}$, while data from ERA5 feature a finer resolution ($0.25^{\circ}\times0.25^{\circ}$). Each source offers \textit{monthly} records for two climate indicators, namely average temperatures (measured in Celsius degrees, $\degree C$) and total precipitations (in millimeters, mm), with the exception of CSIC, which provides monthly records for a third climate variable, the Standardized Precipitation-Evapotranspiration Index\cite{vicente2010multiscalar}, also known as SPEI (unit free). In addition to monthly data, ERA5 also provides records at the temporal resolution of \textit{hours}, which we aggregate to obtain \textit{daily} values.

CRU TS employs raw data from an extensive network of weather stations, computes monthly climate anomalies, and interpolates them using angular-distance weighting\cite{harris2020version} (ADW). ADW is employed to account for the varying area represented by each grid cell on a spherical Earth, in particular by considering the cosine of the latitude of each grid cell. The cosine of the latitude serves as a measure of the change in grid cell area with respect to latitude. Cells near the equator have larger areas as compared to those near the poles, where cells are smaller.

CSIC leverages CRU TS data to provide the SPEI, a drought index that combines information from both precipitation and evapotranspiration to assess the severity and duration of drought conditions. It is a standardized version of the widely used Palmer Drought Severity Index (PDSI) that takes into account the effects of both precipitation and temperature on water availability. Given its multi-scalar nature, it is able to differentiate among different types of drought; we currently propose the 1-month level of aggregation, focusing on changes in headwater levels.

ERA5 climate data set uses data from radiosondes, which are battery-powered telemetry instruments carried into the atmosphere by weather balloons to measure various atmospheric parameters, including temperature, wind, and humidity profiles. The information collected by radiosondes is transmitted back to the ground via radio signals and is assimilated by ERA5 along with other observations, such as satellite and surface-based measurements, to provide a comprehensive picture of the Earth's climate system\cite{hersbach2020era5}.

Finally, UDEL provides gridded estimates mainly based on station records compiled from several publicly available sources (e.g., Global Historical Climatology Network dataset \cite{peterson1997overview}, Global Historical Climatology Network Monthly dataset \cite{lawrimore2011overview}, the Daily Global Historical Climatology Network archive \cite{menne2012overview}). Interpolation is performed with Shepard spatial-interpolation algorithm\cite{shepard1968two}, modified for use over Earth’s near-spherical surface.

\begin{table}[h!]
    \centering
    \caption{Summary of the main features of the employed data sources}
    \label{tab:sources}
    \begin{tabular}{llllll}
    \toprule
    \textbf{Source} & 
    Reanalysis\footnotemark
      & \textbf{Variables} & \textbf{Coverage period} & \textbf{Resolution} & \textbf{Version} \\
    \midrule
    CRU TS \cite{harris2020version} & No & Temperature, precipitation & 1901--2022 & $0.5^{\circ}$ & 4.07 \\
    CSIC \cite{vicente2010new} & No & SPEI 1-month & 1901--2020 & $0.5^{\circ}$ & 2.7\\
    ERA5 \cite{hersbach2020era5} & Yes & Temperature, precipitation & 1940--2023 & $0.25^{\circ}$ & 5\\
    UDEL \cite{willmott2000terrestrial} & No & Temperature, precipitation & 1900--2017 & $0.5^{\circ}$ & 5.01  
    \\ \midrule
    NASA \cite{doxsey2015taking} & & Population density & 2000, 2005, 2010, 2015 & $0.25^{\circ}$ & 4 \\
    Li et al. (2020) \cite{li2020harmonized} &  & Night-light intensity & 2000, 2005, 2010, 2015 & $0.008\Bar{3}^{\circ}$ & 7\\ \midrule
    GADM \cite{gadm} & & Administrative boundaries & & & 4.1 \\
    \bottomrule
    \end{tabular}
\end{table}
\footnotetext[1]{\textit{Reanalysis} refers to the integration of climate models with past observations to provide (i) consistent values over time and (ii) more accurate estimates in the grids not covered by measurement stations.}

\subsubsection*{Socio-economic data}
We use gridded socio-economic data to gauge information on the spatial distribution of economic and human-based activities. In particular, two distinct indicators are used as weights for the spatial aggregation of climate data into administrative units. The first proxy is population density, available from Columbia University's Gridded Population of the World v4 \cite{doxsey2015taking}, measured at $0.25^{\circ}$ and $0.5^{\circ}$ spatial resolutions. The climate econometrics literature has largely employed population density as an indicator of economic activity proxying local exposure to weather conditions \cite{burke2015global,hsiang2016climate,auffhammer2018quantifying}. Note that population density is measured with respect to the land area of each grid. Thus, in our aggregation strategy, we employ the product between the population density and the area of the associated grid to account for population size properly. 

\begin{figure}[h!]
    \centering
    \includegraphics[width=\textwidth]{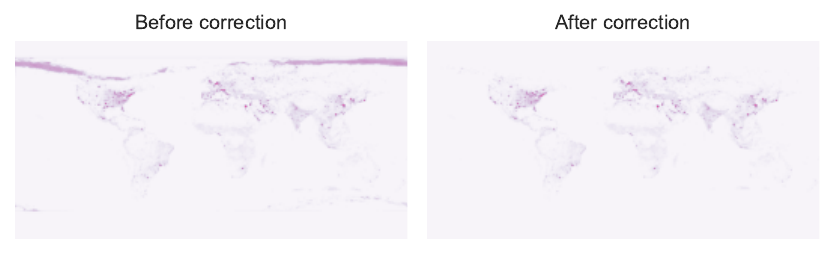}
    \caption{Correction of auroras in the night lights data for the year 2015. The left plot shows night light data \textit{before} correction; the right plot shows the same data \textit{after} correction, which consists of setting to 0 the radiance linked to grids above the 45 North and below the 45 South parallels in case the value of the grids were equal to 0 in the night light data of 2000, 2005, and 2010, which were not affected by the auroras issue.}
    \label{fig:aurora}
\end{figure}

A second, alternative indicator of economic activity that we include in our repository is night-time light data \cite{li2020harmonized}, which is originally available at a 30 arc-second spatial resolution ($0.008\Bar{3}^{\circ}$). To match this finer resolution with the coarser resolutions of our gridded climate data, we compute the mean of the values of the cells in the $0.25^{\circ}$ and $0.5^{\circ}$ grids. 
We aggregate by first taking the mean of 900 (30$\times$30) and 3600 (60$\times$60) most upper-left cells in our coordinate system to produce a single grid at a resolution of, respectively, $0.25^{\circ}$ and $0.5^{\circ}$. We then iterate this procedure with the adjacent blocks of cells to obtain all the gridded values of the night-light data for the coarser resolution. We note that the harmonized VIIRS-DMSP \texttt{tif} file for the year 2015 presented noise from auroras, especially in the northern hemisphere (see Figure \ref{fig:aurora}, left panel). Therefore, following a standard procedure \cite{li2020harmonized}, we set to 0 the radiance linked to grids above the 45 North and below the 45 South parallels when they had null values in 2000, 2005, and 2010. Figure \ref{fig:aurora}, right panel, shows the result of this correction.

We allow weighting by either population or night-lights using the base years 2000, 2005, 2010, and 2015\footnote{The code needed to replicate analysis with additional base years is available in the repository associated with the paper.}. Moreover, the repository allows to explore and/or download climate data without weighting them by any spatial economic indicator. This option is referred to as \enquote{unweighted}.

\subsubsection*{Administrative boundaries}
We employ two levels of geographical resolution from the Database of Global Administrative Areas \cite{gadm} (GADM). While the first level (GADM0) has a coarser resolution and replicates country boundaries, the second level (GADM1) is sub-national and consists of the largest administrative area included within national countries (e.g., states for the US, regions for Italy, etc.). In our work, we used GADM version 4.1 released on July 16, 2022.

\subsection*{Weighting and aggregation strategy}
Raw grid data require to be aggregated to match administrative areas, for which many other socioeconomic indicators are usually available. The general weighting scheme is the following:
\begin{equation}
    \label{eq:aggregation}
    y_{i,t,w,T} = \frac{\sum_{j \in J_i} a_{j} f_{i,j} w_{j,T} x_{i,t}} {\sum_{j \in J_i} a_{j} f_{i,j} w_{j,T}}
\end{equation}
where $y_{i,t,w,T}$ is the value of the climate variable $y$ in the geographical unit $i$ (at a specified GADM resolution) at time $t$ weighted by proxy $w$ measured in base year $T\in\{2000,2005,2010,2015\}$; $J_i$ is the set of grids intersecting the geographic unit $i$; $f_{i,j}$ is the fraction of grid $j$ which intersects the geographic unit $i$; $a_{j}$ is the area of the grid $j$; $x_{i,t}$ is the raw grid climate variable. In line with the prevailing practice in the literature, the base year $T$ is fixed ex-ante and does not vary with $t$ \cite{burke2015global,auffhammer2018quantifying}. Of course, for the unweighted aggregation, $w_{j,T}=1$ for any $j$ and $T$.

\begin{figure}[h!]
    \centering
    \includegraphics[width=0.5\textwidth]{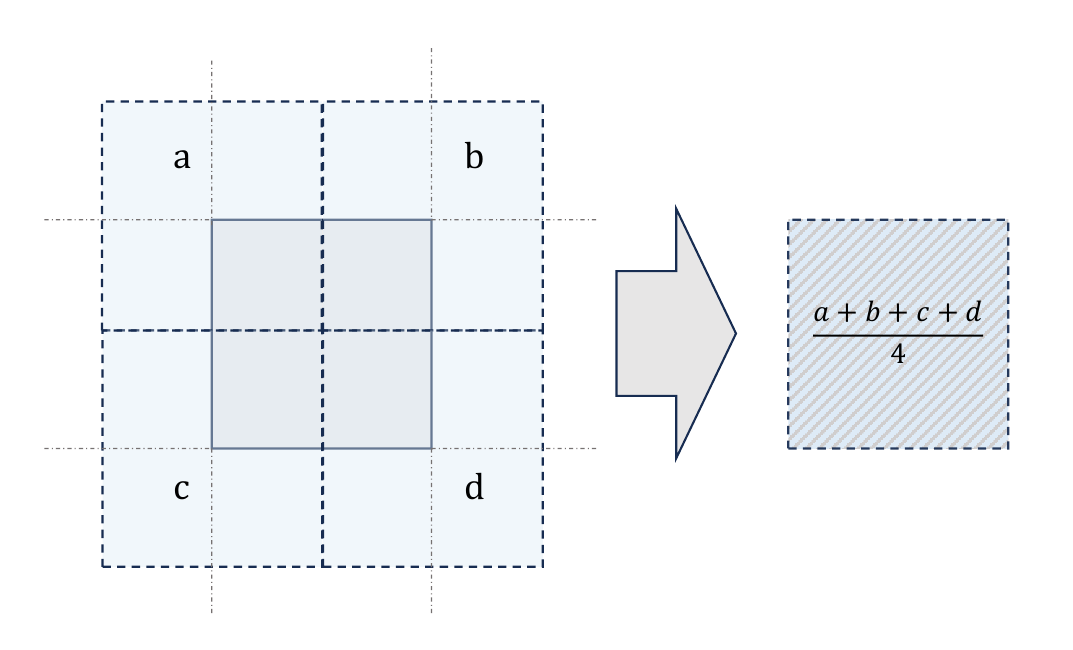}
    \caption{Stylized illustration of the bilinear interpolation when the population density and night-time lights grids are used to weigh ERA5 climate variables. The extent of the weighting grids slightly differs from the extent of the ERA5 climate variables grids, both in longitude and latitude, resulting in a difference of $0.125^{\circ}$ in both directions, exactly half of the spatial resolution of the ERA5 data. Since our weighting procedure requires weighting and variable grids to overlap, we resample the weighting grids, filling the values with a simple average of the values of the intersecting grids.}
    \label{fig:resample_ERA5}
\end{figure}

We notice that grid resolutions may vary across data sources. The \texttt{NetCDF} files retrievable from ERA5 is made up of a 721$\times$1440 grid, with extremities (180.125$^{\circ}$W, 179.875$^{\circ}$E, 90.125$^{\circ}$S, 90.125$^{\circ}$N), with a 15 arc-minute spatial resolution. The gridded file of the population density and night-time lights feature instead a 720$\times$1440 grid, with extremities (180$^{\circ}$W, 180$^{\circ}$E, 90$^{\circ}$S, 90$^{\circ}$N). To make weighting and climate variables from ERA5 consistent with population data, we resampled the values of the weight grids with a simple bilinear interpolation. The logic behind such a procedure is sketched in Figure \ref{fig:resample_ERA5}, where the stylized grids of two sources are displayed. This procedure is applied whenever we weigh climate variables from ERA5 with population density and night-time light grid files.

\begin{figure}[h!]
    \centering
    \includegraphics[width=\textwidth]{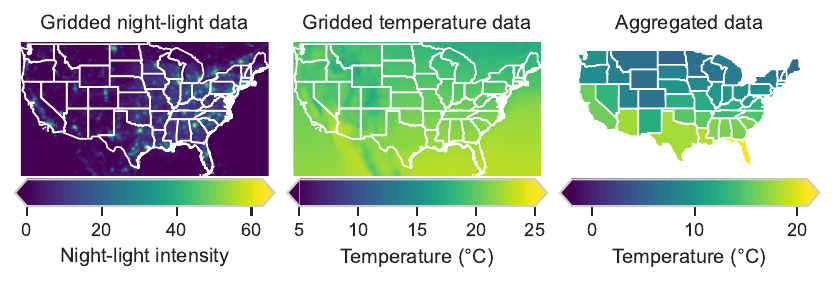}
    \caption{Example of climate data weighting for the US. The left panel shows raw gridded night-light data in 2015. The middle panel displays raw gridded temperature data in 2015. Finally, the right panel shows, for the year 2015, temperatures aggregated at the GADM1 administrative level \textit{weighted by} night lights in 2015. 
    }
    \label{fig:US_eg}
\end{figure}

As an example of the aggregation strategy, Figure \ref{fig:US_eg} shows three panels. The left and center panels display raw gridded data for night-light intensity in 2015, and ERA5 average annual temperatures in 2015 for the contiguous US, respectively. Nightlights are chosen as the weighting variable in this example. Figure \ref{fig:US_eg}, right panel, displays the resulting aggregation at GADM1 resolution and illustrates the output that users can retrieve from our repository. 

\subsection*{Customization}
Once aggregated, data can be customized in different ways. First, users can select a time interval by defining specific initial and final time boundaries. Second, it is possible to choose among different time frequencies. As mentioned before, daily data are only provided by ERA5. Notice also that weighted values for daily and monthly observations are computed using raw data at the same temporal resolution from the original sources. Annual observations are instead obtained by aggregating monthly to annual observations, via the computation of relevant summary statistics (i.e., average for temperatures, sum for precipitations), without seasonal adjustments. The only exception is the SPEI variable, for which daily data are not available and for which a simple average is not meaningful (i.e., SPEI cannot be linearly aggregated). As a result, we provide SPEI only at a monthly temporal resolution. Third, exploiting daily data provided by ERA5, the repository provides information about the frequency of extreme weather conditions. In particular, users can specify either an absolute or a relative threshold for the climate variable of interest. The repository then returns the number of days for which the climate variable has attained values over that threshold, for each geographical unit, month or year. For example, if the users sets for temperature an absolute threshold of $20^\circ C$, the web app returns the number of days for which the average temperature of the geographical unit has exceeded the threshold within the chosen month or year. Similarly, users can set a relative (quantile) threshold of $0.90$. In this case, the web app returns the number of days for which the average temperature of the geographical unit has fallen in the top 10\% percentile within the chosen month or year. For each geographical unit, the percentile is computed on its historical distribution, i.e. the distributions of temperatures in that region or country \textit{regardless} time measurement.

\section*{Data Records}
Our repository contains 138 data sets, each referring to a specific combination of geographical resolution (GADM0, GADM1), climate variable (temperature, precipitation, SPEI), climate data source (CRU TS, UDEL, ERA5, CSIC), weighting variable (unweighted, population density, night-lights), time resolution (daily, monthly), and weighting base year (2000, 2005, 2010, 2015). Data can be downloaded from \url{https://weightedclimatedata.streamlit.app} in two different formats: the wide format has geographical units as keys and values of a climate variable in different years as attributes; the long format has geographical units and years as keys, and the value of a climate variable as only attribute. Data can also be downloaded in three different extensions (\texttt{csv}, \texttt{json}, and \texttt{parquet}). 

\begin{figure}[h!]
    \centering
    \includegraphics[width=0.88\textwidth]{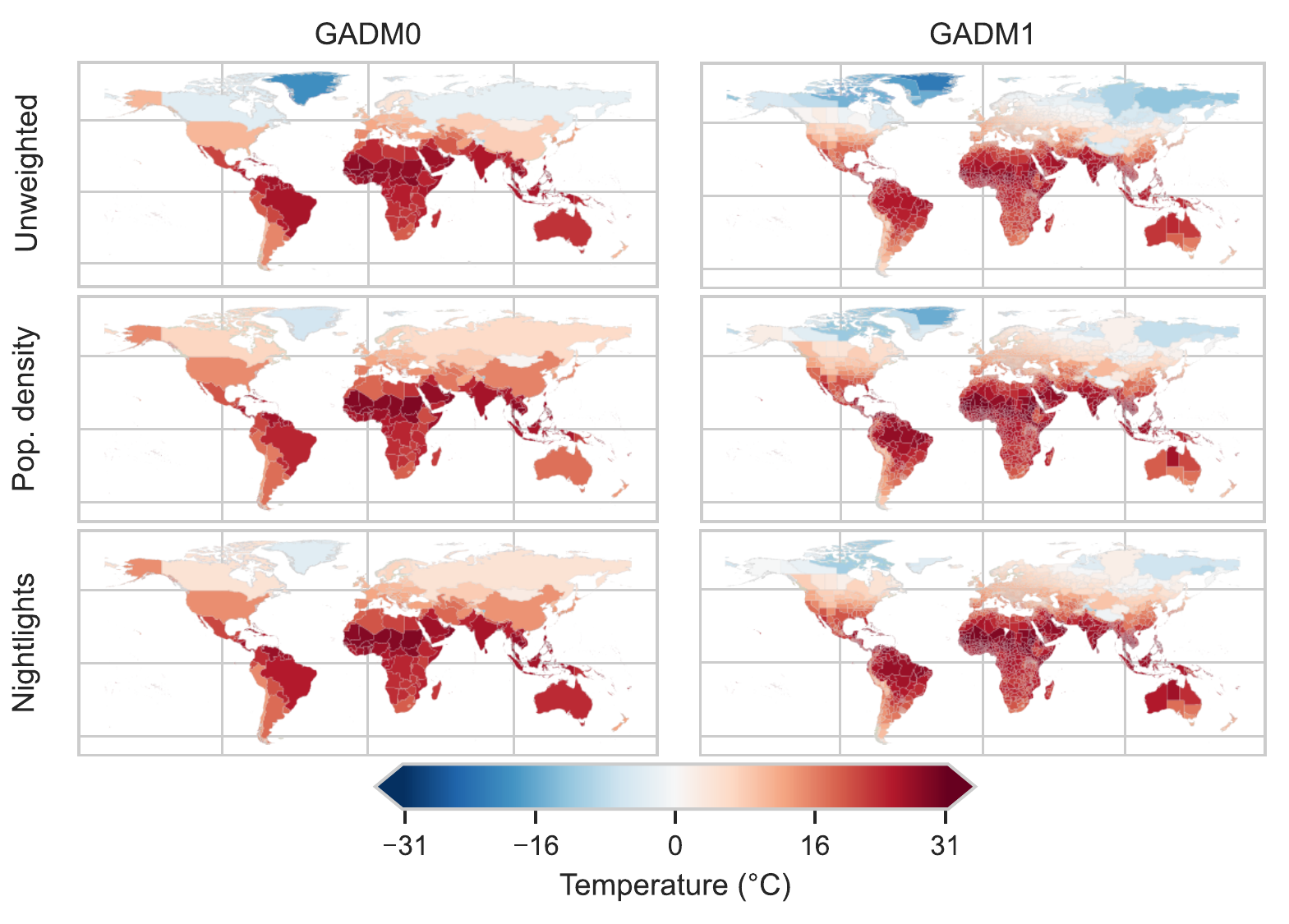}
    \caption{Comparison among different weighting and aggregation schemes; ERA5, temperatures in 2015, weighting base year 2015.}
    \label{fig:gadm01}
\end{figure}

\begin{figure}[h!]
    \centering
    \includegraphics[width=0.88\textwidth]{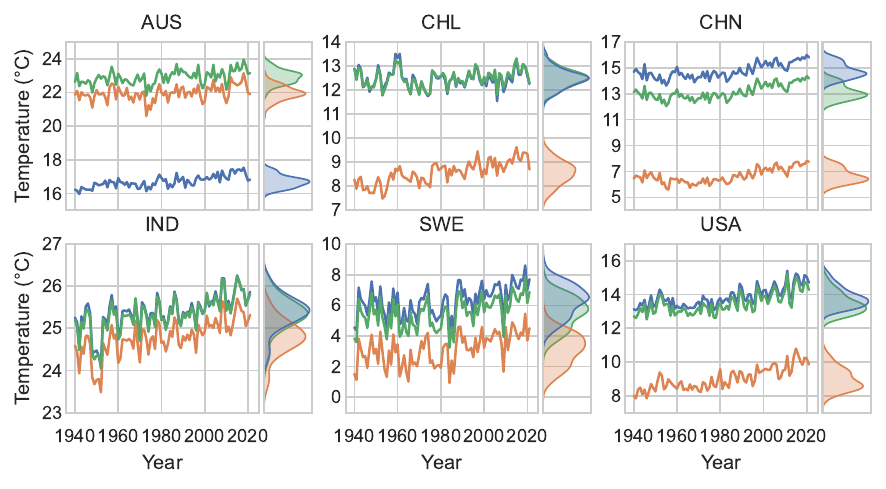}
    \caption{Time series of temperatures from ERA5 for selected countries (from top-left to bottom-right: Australia, Chile, China, India, Sweden, USA), weighted by population density (blue), night lights (green) in 2015, or unweighted (orange).}
    \label{fig:distance_ts}
\end{figure}

The results obtained from the pipelines described in the previous section are exemplified and illustrated in Figures \ref{fig:gadm01} and \ref{fig:distance_ts}. Figure \ref{fig:gadm01} shows World maps at two different geographical resolution levels: GADM0 (country) and GADM1 (sub-national). These maps illustrate the ERA5 temperature levels in year 2015, considering various weighting variables. The maps provide a visual representation of how temperature values vary across countries and regions after being weighted by different indicators of economic activity. It is immediately visible how weighting for proxies of local anthropogenic activity returns a different picture of cross-country differences in experienced average temperature as compared to unweighted data, especially for areas at extreme latitudes. Further, while population and night-light weights provide relatively similar average temperatures in 2015 at country level, remarkable differences are detectable when scaling at the sub-national level, especially in Latin America, Central and Southern Africa, and Australia, which are areas where economies are among the most exposed to climate damages\cite{burke2015global,palagi2022climate}. 

Taking a different perspective, Figure \ref{fig:distance_ts} shows the time series of ERA5 temperatures weighted by different indicators of local economic activity for six selected countries from 1940 to 2020. The six panels of Figure \ref{fig:distance_ts} demonstrate how temperature values, when aggregated under different weighting schemes, fluctuate over time within each country. While for some countries the choice of the weighting and aggregation scheme is almost irrelevant (e.g., India; bottom-left panel), in others it provides very different average temperature levels (e.g., Australia; top-left panel) and long-run dynamics (e.g., Chile; top-center panel).

\section*{Technical Validation}
In this section, we validate the datasets constructed in our repository against those employed in two influential climate econometric exercises: Kotz et al. \cite{kotz2022effect} and Burke at al.\cite{burke2015global}. We evaluate the agreement between our weighting procedures and those obtained by these two studies, with the aim of supporting the reliability and effectiveness of our approach.

In order to conduct a proper validation exercise, we first align our data sources with the exact versions employed by the two targeted studies, which of course have been employing older versions for both climate and economic activity datasets. This allows us to validate the accuracy and robustness of our data processing pipelines and methods and to ensure a fair and reliable assessment of the quality and consistency of our estimates. 

More precisely, Burke et al.\cite{burke2015global} exploit UDEL v3.01 for precipitation and temperature data, and v3 of the NASA $0.50^{\circ}$ gridded population data in 2000. Population is used as the weighting variable and, although the authors do not specify the source and version of the national administrative boundaries they use, their shape files are publicly available. Conversely, Kotz et al. \cite{kotz2022effect} use $0.25^{\circ}$ gridded ERA5 precipitation and temperature data, do not weigh climate data with any indicator of economic activity, and employ GADM1 v3.6 for the spatial aggregation.

\begin{figure}[h!]
    \centering
    \includegraphics[width=\textwidth]{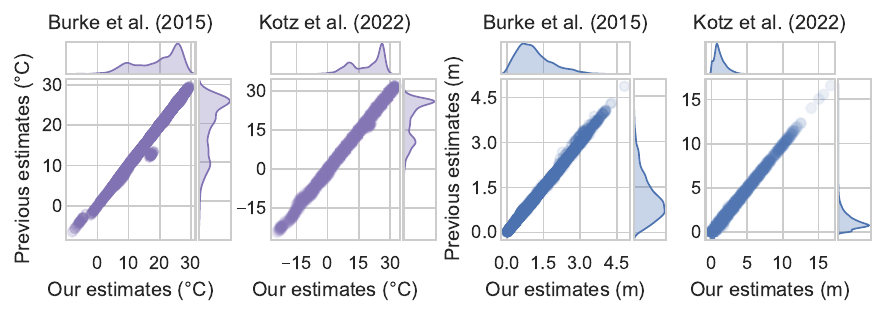}
    \caption{Comparison of weighted and/or aggregated temperature and precipitation variables in our datasets against data used in Burke et al. (countries) and Kotz et al. (sub-national regions). Average yearly temperature is expressed in degrees Celsius while annual total precipitations are in meters. Values on the main diagonal indicate very similar estimates.}
    \label{fig:cmp}
\end{figure}

Results of our comparative analysis are reported in Figure \ref{fig:cmp}. The figure includes four scatterplots, each representing the relationship between our estimates and those used in the original studies for both temperature and rainfall (SPEI is not used in either of the two). Intuitively, points aligning on the main diagonal of the scatterplots indicate agreement and reflect the similarity between the estimates.

It is important to note that the data shown in Figure \ref{fig:cmp} encompass all the years analyzed in the original studies. Notably, a substantial majority of our estimates exhibit a high degree of correspondence with the weighted and/or aggregated data employed by previous authors. This indicates a strong level of agreement between our results and those of the earlier studies, corroborating the quality and reliability of the methods employed in our repository. However, there also emerge some minor discrepancies that are worth pointing out. In particular, the first panel on the left highlights two main sources of disagreement between the estimates of Burke et al. and ours. The first one, on the bottom left (where both temperatures are negative), regards Greenland. In this case, the estimates of Burke et al. are less conservative than ours. The second one, where our estimates are instead slightly more generous than the ones by Burke et al., concerns Bhutan. These discrepancies are mainly due to the weighting scheme, and in particular to the fact that population density is highly concentrated in a few regions of Greenland and Bhutan.

\section*{Usage Notes}
\begin{figure}
    \centering
    \includegraphics[height=0.95\textheight]{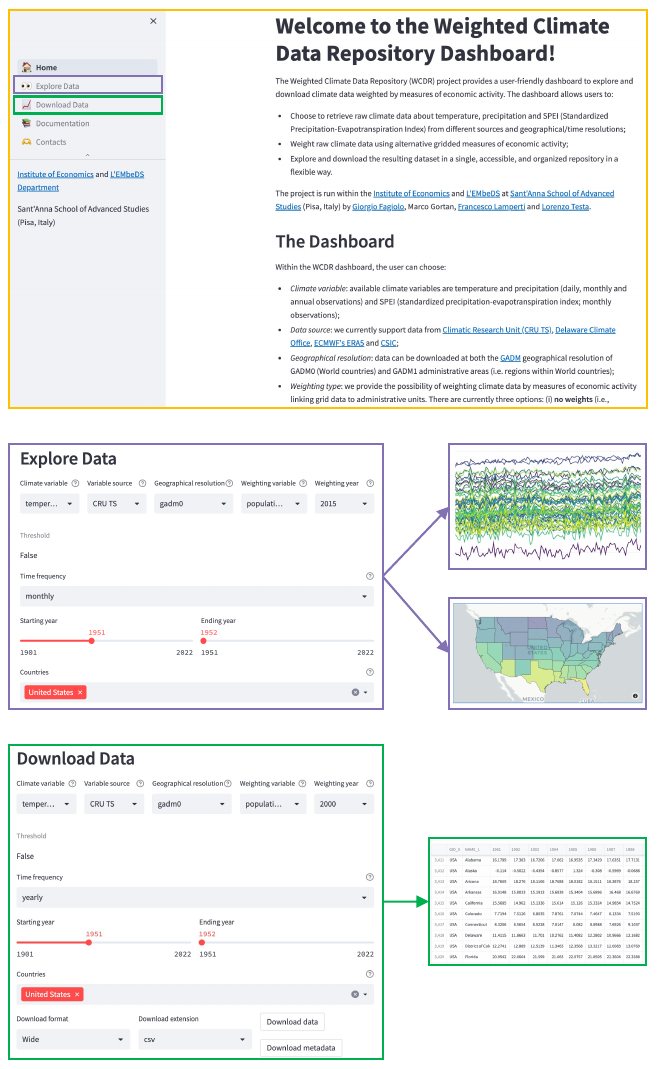}
    \caption{Dashboard interface, examplified}
    \label{fig:dashboard}
\end{figure}

The \textit{Weighted Climate Data Repository} dashboard can be accessed at \url{https://weightedclimatedata.streamlit.app}. The homepage provides an overview of the main features of the web app, introducing users to its visualization and downloading features, which can be experienced by clicking on the respective links on the left-sidebar (see Figure \ref{fig:dashboard}, top panel, for an illustration). 

The visualization tab, shown in Figure \ref{fig:dashboard} (central panel), provides an easy interface with plotting tools. In particular, users can choose among different options (i.e. climate variables, variable sources, geographical resolutions, weighting indicators, weight base years, threshold options, time frequencies) and filters (i.e. starting and ending years, observations) to produce two kinds of plots in real-time: time series and choropleth maps. 

Finally, Figure \ref{fig:dashboard} (bottom panel) shows the download tab. Similarly to the visualization tab, users can customize the files they want to download through several options and filters. In addition to the previous tab, users can also select the data format and extension, as discussed above. Before downloading the data, the dashboard provides a preview option, facilitating the overall user experience. Metadata containing source versions can be downloaded as well. 

\section*{Future extensions}
We envision several extensions to our current work. By keeping the repository updated to the latest versions of the available data providers, we first aim at covering longer time spans. We also plan to include higher spatial resolution geographical observations, such as GADM2, enhancing the spatial detail of data offered by our repository. Furthermore, we aim to introduce more climate variables and more upscaling summary statistics, enriching the range of possible derived climate indicators generated by the repository. Under this respect, we also plan to extend also the SPEI climate variable to other scales, up until the 48-month variant, which is the longest available.

\section*{Code availability}

\texttt{Python} code running the \textit{Weighted Climate Data Repository} dashboard and scripts for aggregating data are available at \url{https://github.com/CoMoS-SA/climaterepo}. 
The \textit{Weighted Climate Data Repository} leverages \texttt{Streamlit}.
We employed \texttt{R} to process the data, exploiting package \texttt{exactextract} (version 0.9.1) for the weighted aggregations.

\bibliography{main}

\begin{thebibliography}{10}
\urlstyle{rm}
\expandafter\ifx\csname url\endcsname\relax
  \def\url#1{\texttt{#1}}\fi
\expandafter\ifx\csname urlprefix\endcsname\relax\def\urlprefix{URL }\fi
\expandafter\ifx\csname doiprefix\endcsname\relax\def\doiprefix{DOI: }\fi
\providecommand{\bibinfo}[2]{#2}
\providecommand{\eprint}[2][]{\url{#2}}

\bibitem{dell2014we}
\bibinfo{author}{Dell, M.}, \bibinfo{author}{Jones, B.~F.} \& \bibinfo{author}{Olken, B.~A.}
\newblock \bibinfo{journal}{\bibinfo{title}{What do we learn from the weather? {T}he new climate-economy literature}}.
\newblock {\emph{\JournalTitle{Journal of Economic literature}}} \textbf{\bibinfo{volume}{52}}, \bibinfo{pages}{740--798} (\bibinfo{year}{2014}).

\bibitem{carleton2016social}
\bibinfo{author}{Carleton, T.~A.} \& \bibinfo{author}{Hsiang, S.~M.}
\newblock \bibinfo{journal}{\bibinfo{title}{Social and economic impacts of climate}}.
\newblock {\emph{\JournalTitle{Science}}} \textbf{\bibinfo{volume}{353}}, \bibinfo{pages}{aad9837} (\bibinfo{year}{2016}).

\bibitem{schlenker2009nonlinear}
\bibinfo{author}{Schlenker, W.} \& \bibinfo{author}{Roberts, M.~J.}
\newblock \bibinfo{journal}{\bibinfo{title}{Nonlinear temperature effects indicate severe damages to {US} crop yields under climate change}}.
\newblock {\emph{\JournalTitle{Proceedings of the National Academy of sciences}}} \textbf{\bibinfo{volume}{106}}, \bibinfo{pages}{15594--15598} (\bibinfo{year}{2009}).

\bibitem{burke2015global}
\bibinfo{author}{Burke, M.}, \bibinfo{author}{Hsiang, S.~M.} \& \bibinfo{author}{Miguel, E.}
\newblock \bibinfo{journal}{\bibinfo{title}{Global non-linear effect of temperature on economic production}}.
\newblock {\emph{\JournalTitle{Nature}}} \textbf{\bibinfo{volume}{527}}, \bibinfo{pages}{235--239} (\bibinfo{year}{2015}).

\bibitem{kalkuhl2020impact}
\bibinfo{author}{Kalkuhl, M.} \& \bibinfo{author}{Wenz, L.}
\newblock \bibinfo{journal}{\bibinfo{title}{The impact of climate conditions on economic production. {E}vidence from a global panel of regions}}.
\newblock {\emph{\JournalTitle{Journal of Environmental Economics and Management}}} \textbf{\bibinfo{volume}{103}}, \bibinfo{pages}{102360} (\bibinfo{year}{2020}).

\bibitem{kotz2022effect}
\bibinfo{author}{Kotz, M.}, \bibinfo{author}{Levermann, A.} \& \bibinfo{author}{Wenz, L.}
\newblock \bibinfo{journal}{\bibinfo{title}{The effect of rainfall changes on economic production}}.
\newblock {\emph{\JournalTitle{Nature}}} \textbf{\bibinfo{volume}{601}}, \bibinfo{pages}{223--227} (\bibinfo{year}{2022}).

\bibitem{abel2019climate}
\bibinfo{author}{Abel, G.~J.}, \bibinfo{author}{Brottrager, M.}, \bibinfo{author}{Cuaresma, J.~C.} \& \bibinfo{author}{Muttarak, R.}
\newblock \bibinfo{journal}{\bibinfo{title}{Climate, conflict and forced migration}}.
\newblock {\emph{\JournalTitle{Global environmental change}}} \textbf{\bibinfo{volume}{54}}, \bibinfo{pages}{239--249} (\bibinfo{year}{2019}).

\bibitem{palagi2022climate}
\bibinfo{author}{Palagi, E.}, \bibinfo{author}{Coronese, M.}, \bibinfo{author}{Lamperti, F.} \& \bibinfo{author}{Roventini, A.}
\newblock \bibinfo{journal}{\bibinfo{title}{Climate change and the nonlinear impact of precipitation anomalies on income inequality}}.
\newblock {\emph{\JournalTitle{Proceedings of the National Academy of Sciences}}} \textbf{\bibinfo{volume}{119}}, \bibinfo{pages}{e2203595119} (\bibinfo{year}{2022}).

\bibitem{carleton2022}
\bibinfo{author}{Carleton, T.} \emph{et~al.}
\newblock \bibinfo{journal}{\bibinfo{title}{Valuing the global mortality consequences of climate change accounting for adaptation costs and benefits}}.
\newblock {\emph{\JournalTitle{The Quarterly Journal of Economics}}} \textbf{\bibinfo{volume}{137}}, \bibinfo{pages}{2037--2105} (\bibinfo{year}{2022}).

\bibitem{auffhammer2014measuring}
\bibinfo{author}{Auffhammer, M.} \& \bibinfo{author}{Mansur, E.~T.}
\newblock \bibinfo{journal}{\bibinfo{title}{Measuring climatic impacts on energy consumption: A review of the empirical literature}}.
\newblock {\emph{\JournalTitle{Energy Economics}}} \textbf{\bibinfo{volume}{46}}, \bibinfo{pages}{522--530} (\bibinfo{year}{2014}).

\bibitem{hsiang2016climate}
\bibinfo{author}{Hsiang, S.}
\newblock \bibinfo{journal}{\bibinfo{title}{Climate econometrics}}.
\newblock {\emph{\JournalTitle{Annual Review of Resource Economics}}} \textbf{\bibinfo{volume}{8}}, \bibinfo{pages}{43--75} (\bibinfo{year}{2016}).

\bibitem{auffhammer2018quantifying}
\bibinfo{author}{Auffhammer, M.}
\newblock \bibinfo{journal}{\bibinfo{title}{Quantifying economic damages from climate change}}.
\newblock {\emph{\JournalTitle{Journal of Economic Perspectives}}} \textbf{\bibinfo{volume}{32}}, \bibinfo{pages}{33--52} (\bibinfo{year}{2018}).

\bibitem{diffenbaugh2019global}
\bibinfo{author}{Diffenbaugh, N.~S.} \& \bibinfo{author}{Burke, M.}
\newblock \bibinfo{journal}{\bibinfo{title}{Global warming has increased global economic inequality}}.
\newblock {\emph{\JournalTitle{Proceedings of the National Academy of Sciences}}} \textbf{\bibinfo{volume}{116}}, \bibinfo{pages}{9808--9813} (\bibinfo{year}{2019}).

\bibitem{alessandri2021macroeconomic}
\bibinfo{author}{Alessandri, P.} \& \bibinfo{author}{Mumtaz, H.}
\newblock \bibinfo{journal}{\bibinfo{title}{The macroeconomic cost of climate volatility}}.
\newblock {\emph{\JournalTitle{arXiv preprint arXiv:2108.01617}}}  (\bibinfo{year}{2021}).

\bibitem{spinoni2021global}
\bibinfo{author}{Spinoni, J.} \emph{et~al.}
\newblock \bibinfo{journal}{\bibinfo{title}{Global population-weighted degree-day projections for a combination of climate and socio-economic scenarios}}.
\newblock {\emph{\JournalTitle{International Journal of Climatology}}} \textbf{\bibinfo{volume}{41}}, \bibinfo{pages}{5447--5464} (\bibinfo{year}{2021}).

\bibitem{wei2022comparison}
\bibinfo{author}{Wei, R.}, \bibinfo{author}{Li, Y.}, \bibinfo{author}{Yin, J.} \& \bibinfo{author}{Ma, X.}
\newblock \bibinfo{journal}{\bibinfo{title}{Comparison of weighted/unweighted and interpolated grid data at regional and global scales}}.
\newblock {\emph{\JournalTitle{Atmosphere}}} \textbf{\bibinfo{volume}{13}}, \bibinfo{pages}{2071} (\bibinfo{year}{2022}).

\bibitem{ponticelli2023temperature}
\bibinfo{author}{Ponticelli, J.}, \bibinfo{author}{Xu, Q.} \& \bibinfo{author}{Zeume, S.}
\newblock \bibinfo{title}{Temperature and local industry concentration}.
\newblock \bibinfo{type}{Tech. Rep.}, \bibinfo{institution}{National Bureau of Economic Research} (\bibinfo{year}{2023}).

\bibitem{donadelli2022temperature}
\bibinfo{author}{Donadelli, M.}, \bibinfo{author}{J{\"u}ppner, M.} \& \bibinfo{author}{Vergalli, S.}
\newblock \bibinfo{journal}{\bibinfo{title}{Temperature variability and the macroeconomy: A world tour}}.
\newblock {\emph{\JournalTitle{Environmental and Resource Economics}}} \textbf{\bibinfo{volume}{83}}, \bibinfo{pages}{221--259} (\bibinfo{year}{2022}).

\bibitem{cipollini2023temperature}
\bibinfo{author}{Cipollini, A.}, \bibinfo{author}{Parla, F.} \emph{et~al.}
\newblock \bibinfo{journal}{\bibinfo{title}{Temperature and growth: A panel mixed frequency {VAR} analysis using {NUTS2} data}}.
\newblock {\emph{\JournalTitle{RECENT WORKING PAPER SERIES}}}  (\bibinfo{year}{2023}).

\bibitem{donadelli2021global}
\bibinfo{author}{Donadelli, M.}, \bibinfo{author}{Gr{\"u}ning, P.}, \bibinfo{author}{J{\"u}ppner, M.} \& \bibinfo{author}{Kizys, R.}
\newblock \bibinfo{journal}{\bibinfo{title}{Global temperature, {R\&D} expenditure, and growth}}.
\newblock {\emph{\JournalTitle{Energy Economics}}} \textbf{\bibinfo{volume}{104}}, \bibinfo{pages}{105608} (\bibinfo{year}{2021}).

\bibitem{harris2020version}
\bibinfo{author}{Harris, I.}, \bibinfo{author}{Osborn, T.~J.}, \bibinfo{author}{Jones, P.} \& \bibinfo{author}{Lister, D.}
\newblock \bibinfo{journal}{\bibinfo{title}{Version 4 of the {CRU TS} monthly high-resolution gridded multivariate climate dataset}}.
\newblock {\emph{\JournalTitle{Scientific data}}} \textbf{\bibinfo{volume}{7}}, \bibinfo{pages}{109} (\bibinfo{year}{2020}).

\bibitem{vicente2010new}
\bibinfo{author}{Vicente-Serrano, S.~M.}, \bibinfo{author}{Beguer{\'\i}a, S.}, \bibinfo{author}{L{\'o}pez-Moreno, J.~I.}, \bibinfo{author}{Angulo, M.} \& \bibinfo{author}{El~Kenawy, A.}
\newblock \bibinfo{journal}{\bibinfo{title}{A new global 0.5 gridded dataset (1901--2006) of a multiscalar drought index: Comparison with current drought index datasets based on the palmer drought severity index}}.
\newblock {\emph{\JournalTitle{Journal of Hydrometeorology}}} \textbf{\bibinfo{volume}{11}}, \bibinfo{pages}{1033--1043} (\bibinfo{year}{2010}).

\bibitem{hersbach2020era5}
\bibinfo{author}{Hersbach, H.} \emph{et~al.}
\newblock \bibinfo{journal}{\bibinfo{title}{The {ERA5} global reanalysis}}.
\newblock {\emph{\JournalTitle{Quarterly Journal of the Royal Meteorological Society}}} \textbf{\bibinfo{volume}{146}}, \bibinfo{pages}{1999--2049} (\bibinfo{year}{2020}).

\bibitem{willmott2000terrestrial}
\bibinfo{author}{Willmott, C.~J.} \& \bibinfo{author}{Matsuura, K.}
\newblock \bibinfo{journal}{\bibinfo{title}{Terrestrial air temperature and precipitation: Monthly and annual time series (1950-1996)}}.
\newblock {\emph{\JournalTitle{Center for Climatic Research, Department of Geography, University of Delaware}}}  (\bibinfo{year}{2000}).

\bibitem{vicente2010multiscalar}
\bibinfo{author}{Vicente-Serrano, S.~M.}, \bibinfo{author}{Beguer{\'\i}a, S.} \& \bibinfo{author}{L{\'o}pez-Moreno, J.~I.}
\newblock \bibinfo{journal}{\bibinfo{title}{A multiscalar drought index sensitive to global warming: The standardized precipitation evapotranspiration index}}.
\newblock {\emph{\JournalTitle{Journal of Climate}}} \textbf{\bibinfo{volume}{23}}, \bibinfo{pages}{1696--1718} (\bibinfo{year}{2010}).

\bibitem{peterson1997overview}
\bibinfo{author}{Peterson, T.~C.} \& \bibinfo{author}{Vose, R.~S.}
\newblock \bibinfo{journal}{\bibinfo{title}{An overview of the global historical climatology network temperature database}}.
\newblock {\emph{\JournalTitle{Bulletin of the American Meteorological Society}}} \textbf{\bibinfo{volume}{78}}, \bibinfo{pages}{2837--2850} (\bibinfo{year}{1997}).

\bibitem{lawrimore2011overview}
\bibinfo{author}{Lawrimore, J.~H.} \emph{et~al.}
\newblock \bibinfo{journal}{\bibinfo{title}{An overview of the global historical climatology network monthly mean temperature data set, version 3}}.
\newblock {\emph{\JournalTitle{Journal of Geophysical Research: Atmospheres}}} \textbf{\bibinfo{volume}{116}} (\bibinfo{year}{2011}).

\bibitem{menne2012overview}
\bibinfo{author}{Menne, M.~J.}, \bibinfo{author}{Durre, I.}, \bibinfo{author}{Vose, R.~S.}, \bibinfo{author}{Gleason, B.~E.} \& \bibinfo{author}{Houston, T.~G.}
\newblock \bibinfo{journal}{\bibinfo{title}{An overview of the global historical climatology network-daily database}}.
\newblock {\emph{\JournalTitle{Journal of atmospheric and oceanic technology}}} \textbf{\bibinfo{volume}{29}}, \bibinfo{pages}{897--910} (\bibinfo{year}{2012}).

\bibitem{shepard1968two}
\bibinfo{author}{Shepard, D.}
\newblock \bibinfo{title}{A two-dimensional interpolation function for irregularly-spaced data}.
\newblock In \emph{\bibinfo{booktitle}{Proceedings of the 1968 23rd ACM national conference}}, \bibinfo{pages}{517--524} (\bibinfo{year}{1968}).

\bibitem{doxsey2015taking}
\bibinfo{author}{Doxsey-Whitfield, E.} \emph{et~al.}
\newblock \bibinfo{journal}{\bibinfo{title}{Taking advantage of the improved availability of census data: A first look at the gridded population of the world, version 4}}.
\newblock {\emph{\JournalTitle{Papers in Applied Geography}}} \textbf{\bibinfo{volume}{1}}, \bibinfo{pages}{226--234} (\bibinfo{year}{2015}).

\bibitem{li2020harmonized}
\bibinfo{author}{Li, X.}, \bibinfo{author}{Zhou, Y.}, \bibinfo{author}{Zhao, M.} \& \bibinfo{author}{Zhao, X.}
\newblock \bibinfo{journal}{\bibinfo{title}{A harmonized global nighttime light dataset 1992--2018}}.
\newblock {\emph{\JournalTitle{Scientific data}}} \textbf{\bibinfo{volume}{7}}, \bibinfo{pages}{168} (\bibinfo{year}{2020}).

\bibitem{gadm}
\bibinfo{title}{{GADM}}.
\newblock \bibinfo{howpublished}{\url{https://gadm.org/}}.
\newblock \bibinfo{note}{Accessed: 2022-07-16}.

\end{thebibliography}


\section*{Acknowledgements}
We are grateful to Francesca Chiaromonte,  Daniele Colombo and Damiano Di Francesco for useful feedback and suggestions. F.L.~acknowledges support from the Italian Ministry of Research, PRIN 2022 project ``ECLIPTIC''.

\section*{Author contributions statement}
All authors conceived ideas and analysis approaches. M.G.~and L.T.~retrieved and processed data, implemented pipelines, and performed statistical analyses. All authors wrote the manuscript. G.F.~and F.L.~supervised the research.

\section*{Competing interests}
The authors declare that they have no competing interests.

\end{document}